\documentclass[12pt]{article}
\usepackage{amssymb}
\normalbaselines
\catcode `\@=11
\@addtoreset{equation}{section}

\def\section{\@startsection {section}{1}{\z@}{-3.5ex plus -1ex minus
     -.2ex}{2.3ex plus .2ex}{\normalsize\bf}}
\def\subsection{\@startsection{subsection}{2}{\z@}{-3.25ex plus -1ex minus
 -.2ex}{1.5ex plus .2ex}{\normalsize\bf}}

\def\thebibliography#1{\section*{References\markboth
  {REFERENCES}{REFERENCES}}\list
  {[\arabic{enumi}]}{\settowidth\labelwidth{[#1]}\leftmargin\labelwidth
  \advance\leftmargin\labelsep
  \usecounter{enumi}}
  \def\newblock{\hskip .11em plus .33em minus -.07em}
  \sloppy
  \sfcode`\.=1000\relax}

\catcode `\@=12

\newtheorem{Theorem}{Theorem}[section] 
 
 \font\tenscr=rsfs10
 \font\sevenscr=rsfs7 
 \font\fivescr=rsfs5 
 \skewchar\tenscr='177 \skewchar\sevenscr='177 \skewchar\fivescr='177
 \newfam\scrfam
 \textfont\scrfam=\tenscr
 \scriptfont\scrfam=\sevenscr
 \scriptscriptfont\scrfam=\fivescr
 \def\scr{\fam\scrfam}

 \def\scri{{\scr I}}
      \newcommand{\be}{\begin{equation}}
      \newcommand{\ee}{\end{equation}}
      \newcommand{\E}[1]{{\rm e}^{#1}} 
      \newcommand{\kolo}[1]{\vphantom{#1}\stackrel{\circ}{#1}\!\vphantom{#1}}
      \def\Reals{{\mathbb R}}
      \def\Box{\square}

\def\arcsinh{\mathop{\rm arcsinh}\nolimits}
\newcommand{\dx}[1]{{\mbox{\rm d}#1}} 
\newcommand{\rd}{{\rm d}} 

\begin{document}
\vspace*{2.5cm}
\noindent
{\bf TRAUTMAN-BONDI MASS FOR SCALAR FIELD AND GRAVITY}\vspace{1.3cm}\\
\noindent
\hspace*{1in}

\begin{minipage}{13cm}
 Jacek Jezierski\footnote{Partially supported by a grant KBN Nr 2
  P03A 047 15.}  \vspace{0.3cm}\\ Department
    of Mathematical Methods in Physics,  University of Warsaw,
    ul.  Ho\.za 74, 00-682 Warszawa, Poland \\
    E-mail:  jjacekj@fuw.edu.pl
    \end{minipage}

\vspace*{0.5cm}

\begin{abstract}
\noindent
 The energy at null infinity
 is presented with the help of a simple example of a massless scalar field
in Minkowski spacetime. It is also discussed for Einstein gravity.
In particular, various aspects of the loss of the energy in the radiating
regime are shown.
   \end{abstract}
   
\section{\hspace{-4mm}.\hspace{2mm}INTRODUCTION}
It turns out that the case of the massless scalar field in Minkowski
space-time already exhibits all the essential features of the problem at
hand, while avoiding various technicalities which arise when one wishes to
describe Einstein gravity.
In General Relativity we present a new variational formulation on 
hypersurfaces 
which are space-like inside and light-like near future null infinity
$\scri^+$. The formulae, we obtain, correspond to the mass loss formula.

 \section{\hspace{-4mm}.\hspace{2mm}SCALAR FIELD ON A HYPERBOLOID} 
 Let us
consider a scalar field $\phi$ in a flat Minkowski space $M$ with
the metric \be\label{etaonM} \eta_{\mu\nu}{\rm d}x^\mu{\rm
d}x^\nu= \rho^{-2}\left( -\rho^2{\rm d}s^2 +{2{\rm d}s{\rm d}\rho
\over \sqrt{1+\rho^2}} +{{\rm d}\rho^2\over 1+\rho^2} + {\rm
d}\theta^2 +\sin^2\theta {\rm d}\varphi^2 \right) \ee Let us fix
a coordinate chart $(x^\mu)$ on $M$ such that $x^1=\theta$,
$x^2=\varphi$ (spherical angles), $x^3=\rho$ and $x^0=s$, and let us
denote by $\kolo\gamma_{AB}$ a metric on a unit sphere
($\kolo\gamma_{AB}{\rm d}x^A{\rm d}x^B:={\rm d}\theta^2
+\sin^2\theta {\rm d}\varphi^2$).

We shall consider an initial value problem on a hyperboloid
$\Sigma$ \[ \Sigma_\tau:= \left\{ x\in M \;| \;
x^0=\tau=\mbox{const.}  \right\} \] for our scalar field $\phi$
with a density of the Lagrangian (corresponding to the wave
equation) \[ L:= -\frac 12 \sqrt{-\det \eta_{\mu\nu}}
\eta^{\mu\nu}\phi_\mu\phi_\nu = -\frac 12
\rho^{-2}\sin\theta\left[ \rho^2(\phi_{3})^2 -\frac{(\phi_0)^2
}{1+\rho^2} +{2\phi_{3} \phi_{0} \over \sqrt{1+\rho^2}}
+\kolo\gamma^{AB}\phi_{A} \phi_{B} \right] \]

We use the following convention for indices:  Greek indices $\mu,
\nu, \ldots$ run from 0 to 3; $k,l, \ldots$ are coordinates on a
hyperboloid $\Sigma_\tau$ and run from 1 to 3; $A,B,\ldots$ are
coordinates on $S(\tau,\rho)$ and run from 1 to 2, where $
S(\tau,\rho):= \left\{ x\in \Sigma_\tau \;| \;
x^3=\rho=\mbox{const.}  \right\} $.

The Euler-Lagrange equations for scalar field $\phi$ can be described by
the following generating formula (see e.g. \cite{Kij-Tulcz}) 
 \be\label{Lpole}
\delta L= (p^\mu\delta\phi)_{,\mu} = p^\mu_{,\mu}\delta\phi
 + p^\mu\delta\phi_{\mu} \ee 
where $L=L(\phi,\phi_\mu)$ is the Lagrangian density of the theory.
When we integrate (\ref{Lpole}) over any finite three-volume $V\subset\Sigma$,
 the following variational equation
 \[
\delta \int_V L=\int_V (p^0\delta\phi)_{,0} + \int_{\partial V}
p^3 \delta\phi \]
 can be written\footnote{For the simplicity of notation we assumed that $\partial
 V$ is a sphere $S(\tau,\rho)$. In general case instead of $p^3$ should be 
 normal component $p^{\perp}$ of the canonical momentum $p^{\mu}$ (see \cite{Kij-Tulcz}).}. 
  In particular, the definition 
$p^\mu=\frac{\partial L}{\partial\phi_\mu}$ of the canonical
momenta $p^\mu$ enables one to calculate
 the time and radial components of $p^\mu$
  \[
p^0=\frac {\partial L}{\partial \phi_0}=
\rho^{-2}\sin\theta\left(\frac{\phi_0}{1+\rho^2}
-\frac{\phi_{3}}{\sqrt{1+\rho^2}}\right) \] \[ p^3=\frac
{\partial L}{\partial \phi_3}=-\rho^{-2}\sin\theta \left(\frac
1{\sqrt{1+\rho^2}}\phi_{0}+\rho^{2}\phi_{3}\right) \]
 Let us
observe that in general the integral $ \displaystyle\int_V L $ is
not convergent for $V=\Sigma$, if we assume the usual asymptotics of a
typical solution $\phi$ of the wave equation, namely $\phi=O( \rho)$ and
$\phi_{3}=O\left( 1 \right)$.  The same problem with
``infinities'' we meet in $p^0$ and $p^3$.

 Let
$\overline M$ denote the standard conformal completion at
future null infinity $\scri^+$ of $M$, let $\overline\Sigma_\tau$ be the
closure of $\Sigma_\tau$ in $\overline M$, set
$S_\tau:=S(\tau,\rho=0)=\partial\overline\Sigma_\tau=\overline\Sigma_\tau \cap
\scri^+$.  The reader not familiar with the notion of Scri can simply
think of the $S_\tau$'s as ``spheres at infinity'' on the hypersurfaces
$\Sigma_\tau$. ($\scri^+=\left\{ x\in {\overline M} \;| \;\rho=0\right\}$.)

The conformal rescaling of the metric $\eta_{\mu\nu}$
\be\label{ctr} \eta_{\mu\nu} \longrightarrow
g_{\mu\nu}=\Omega^2\eta_{\mu\nu} \ee
enables one to ``renormalize'' $L$ by adding a full divergence 
  \[ {\overline L}:= -\frac12\sqrt{-g}g^{\mu\nu}\psi_\mu\psi_\nu
+\frac1{12}\sqrt{-g}R(g)\psi^2 = \] \be\label{Lren} = -\frac 12
\Omega^2 \sqrt{-\eta}\eta^{\mu\nu} \psi_\mu \psi_\nu + \frac 12
\Omega\sqrt{-\eta} \Box \Omega \psi^2 = L+\frac 12 \partial_\nu
\left( {\sqrt{-\eta}} \eta^{\mu\nu} \Omega_{,\mu} \Omega \psi^2
\right) \ee where we used a new field variable
$\psi:=\Omega^{-1}\phi$ which is a natural one close to the null
infinity.  The generating formula takes the following form with
respect to the new variable $\psi$ \[\delta \int_V {\overline L}
= \int_V \left( \pi^0 \delta\psi \right)_{,0} + \int_{\partial V}
\pi^3 \delta\psi \] and the Euler-Lagrange equations we write
explicitly \[ \pi^0=\frac {\partial \overline L}{\partial
\psi_0}= \rho^{-2}\Omega^2\sin\theta\left(\frac{\psi_0}{1+\rho^2}
-\frac{\psi_{3}}{\sqrt{1+\rho^2}}\right) \] \[ \pi^3=\frac
{\partial \overline L}{\partial \psi_3}=
-\rho^{-2}\Omega^2\sin\theta \left(\frac
1{\sqrt{1+\rho^2}}\psi_{0} +\rho^{2}\psi_{3} \right) \] \[
\pi^A=\frac {\partial \overline L}{\partial \psi_A}= -
\rho^{-2}\Omega^2\sin\theta\kolo\gamma^{AB}\psi_B \] \[
\partial_\mu \pi^\mu= \frac {\partial \overline L}{\partial
\psi}= \frac16 \sqrt{-g}R(g)\psi \] It is easy to check that all
terms are finite at null infinity, provided $\psi=O(1)$,
$\psi_{3}=O( 1)$ and $\rho^{-1}\Omega=O(1)$.  For the special
case $\Omega=\rho$ see \cite{APP}. It is convenient to ``normalize''
$\Omega$ by assuming that
\be\label{0=1}
\lim_{\rho\rightarrow 0^+}\rho^{-1}\Omega =1 
\ee

   From the above Euler-Lagrange equations one can easily obtain the
conformal wave equation 
\be\label{SS3} \Box_g \psi +\frac16 R(g)\psi =0  \ee

Following the usual practice (\cite{BS}, \cite{Kij-Tulcz}), we perform
the Legendre transformation between $\pi^0$ and $\psi_0$. It gives us
the following Hamiltonian density
 \be\label{Hd} H:= \pi^0\psi_0 - {\overline L} =
\rho^{-2}\Omega^2 \frac 12 \sin\theta\left[ \left(
\rho\psi_{,3}\right)^2 + \frac 1{1+\rho^2}(\psi_0)^2
+\kolo\gamma^{AB}\psi_{A} \psi_{B} -\frac16 \rho^{-2}\Omega^2
R(g) \psi^2 \right] \ee
 and the following variational relation
 \begin{equation} - \delta \int_V H = \int_V \left( {\dot
\pi} \delta \psi - {\dot \psi} \delta \pi \right) +
\int_{\partial V} \pi^3 \delta \psi  \label{pipsi}
\end{equation} where here $\pi:=\pi^0$.

Although the density $H$ given by (\ref{Hd}) depends explicitly on the
choice of the conformal factor $\Omega$, it can be easily checked that
when $V$ becomes $\Sigma$ (extending to the future null infinity)
  the numerical value of the Hamiltonian 
\[ {\cal H}:=\int_\Sigma H \]
is independent on the choice of the conformal factor
 provided $\partial_0\Omega=0$. This can be
easily seen from the following relations 
\[ {\dot \pi} \delta
\psi - {\dot \psi} \delta \pi ={\dot p^0} \delta \phi - {\dot
\phi} \delta p^0 \] \[ \pi^3 \delta \psi - p^3 \delta \phi =
\frac12 \delta \sin\theta\Omega\Omega_3\psi^2 \]
More precisely, the conformal factor  $\Omega$ with
the asymptotics on $\scri^+$
\[ \left. \Omega\right|_{\scri^+} =0 \, ,\quad
\left. \partial_3\Omega\right|_{\scri^+} =1 \]
and $\hat\Omega\equiv 1$ (corresponding to the density of the lagrangian $L$)
give the same numerical value of the Hamiltonian $\cal H$ because
$\Omega\Omega_3\psi^2$ vanishes on $\scri^+$.
Moreover, each term in the variational formula
 \begin{equation} - \delta {\cal H} = \int_\Sigma \left( {\dot
\pi} \delta \psi - {\dot \psi} \delta \pi \right) +
\int_{\partial\Sigma} \pi^3 \delta \psi  \label{pipsionH}
\end{equation}
possesses a universal character not depending on the particular choice of
the conformal factor $\Omega$.

However, when we pass from (\ref{pipsi}) to (\ref{pipsionH}), integrating
 the relation (\ref{pipsi}) over
hyperboloid $\Sigma$, we quickly realize that the boundary term
\[ \int_{\partial \Sigma_\tau} \pi^3\delta \psi= \int_{S_\tau}
\sin\theta\dot\psi\delta \psi (\lim_{\rho\rightarrow 0^+}
\rho^{-2}\Omega^2)=-\int_{\partial \Sigma_\tau} \pi\delta\psi \]
does not vanish for the usual asymptotics\footnote{
The difference in the asymptotics of the field at null and spatial infinity 
is the main obstruction which does not allow to consider the boundary data
at the spatial and at the null infinity in the same way.} of the field $\psi$.
One can try to get a closed Hamiltonian system by  assuming
that $\dot\psi\vert_{\partial\Sigma}=0$ and then the energy will
be conserved in time.  But this is not the case we would like to describe.
We want to consider the
situation with any data $\left.\psi\right|_{\scri^+}$.  In this
case we can define Trautman-Bondi energy\footnote{The explicit
explanation, why we are convinced that the name of Trautman \cite{Trautman}
 should be
associated with the notion of the mass in the radiating regime, is given
in \cite{cjm} but for gravity. We extend here this notion for the energy
of a massless scalar field in the radiating regime.}, but it would be no
longer conserved, formally we can treat it as a Hamiltonian
of the opened Hamiltonian system. The variational formula (\ref{pipsi}) 
enables one to define the TB energy together with
its changes in time.  The boundary data depends explicitly
on time and we are able to compare the initial data
with different boundary conditions.  Moreover, if we extend our phase
space by adding a piece of $\scri^+$, 
we are able to describe a usual closed Hamiltonian system, where
Trautman-Bondi energy is the true Hamiltonian of the dynamics \cite{ptjjk}.

Let us express the density of the Hamiltonian in terms of conformally
rescaled phase variables $(\pi,\psi)$
{\footnotesize
  \be\label{defH} H(\pi,\psi):= \frac 12\rho^{-2}\Omega^2
  \sin\theta\left[ \left( \rho\psi_{3}\right)^2 +
  \left(\frac{\pi\rho^2\sqrt{1+\rho^2}}{\Omega^2\sin\theta}+\psi_3\right)^2
  +\kolo\gamma^{AB}\psi_{A} \psi_{B} -\frac16\rho^{-2}\Omega^2
  R(g)\psi^2 \right] \ee}
   The Hamilton equations are the
  following \be\label{eqm1} \dot\psi =
  \frac{\rho^2\pi}{\Omega^2\sin\theta}(1+\rho^2)+
  \psi_3\sqrt{1+\rho^2} \ee \be\label{eqm2}
  \dot\pi=(\pi\sqrt{1+\rho^2})_{,3} + \left[\rho^{-2}\Omega^2
  (1+\rho^2)\sin\theta\psi_3 \right]_{,3}
  +(\rho^{-2}\Omega^2\sin\theta\kolo\gamma^{AB}\psi_B)_{,A}
  +\frac16\rho^{-2}\Omega^2\sin\theta R(g)\psi \ee and they obviously
  correspond to the wave equation (\ref{SS3}).

  The variational formula (\ref{pipsionH}) describes an opened
Hamiltonian system because we are not allowed to
kill the boundary term.  Our Hamiltonian is not conserved in
time\footnote{The conservation law is an obvious feature of the ADM energy
defined on the spacelike hypersurface $\Sigma$ which is stretching to the
spatial infinity.}
\be\label{Hdot} -\partial_0 {\cal H } =
\int_{\partial \Sigma} \pi^3 \dot\psi = \int_{S_\tau} \sin\theta
(\dot\psi)^2 (\lim_{\rho\rightarrow 0^+} \rho^{-2}\Omega^2)=
\int_{S_\tau} \sin\theta (\dot\psi)^2 
\ee
and the last equality holds for $\Omega$ obeying the boundary condition
(\ref{0=1}). 
Formally, the result (\ref{Hdot}) can be obtained from
(\ref{pipsionH}) if we replace variation $\delta$ with $\partial_0$,
but it can be also checked by a direct computation, using
equations (\ref{eqm1}) and (\ref{eqm2}).

\subsection{\hspace{-5mm}.\hspace{2mm}Trautman-Bondi Mass as a Hamiltonian}

For $\tau > \tau_0$ let $N_{[\tau_0,\tau]}:=\cup_{u\in[\tau_0,\tau]} S_u$,
so $N_{[\tau_0,\tau]}$ is a null hypersurface contained
in $\scri^+$ with boundary $S_\tau\cup S_{\tau_0}$.

An attempt to treat
separately the hyperboloid and Scri leads to the various difficulties in
the Hamiltonian  approach. In particular, we have learned from \cite{APP} that
there is no possibility to get a nice Hamiltonian system for the
hyperboloidal foliation. However, the ADM energy assigned to the
hyperboloid $\Sigma_{\tau}$ plus $N_{]-\infty,\tau]}$ -- a piece of Scri
between hyperboloid and spatial infinity (cf. \cite{APP}), 
enables one to remove an infinite tail $N_{]-\infty,\tau_0]}$ and apply
the remaining Trautman-Bondi energy as a Hamiltonian.
More precisely, let us
consider a Hamiltonian system on a surface $\Sigma_\tau\cup
N_{[\tau,\tau_0]}$.
We propose the following variational formula, which is a direct consequence
of the considerations\footnote{In
the paper \cite{APP} they were applied for the ADM energy.}
 given in subsection 2.4 of \cite{APP}
\footnotesize
 \[ -\delta\left( \int_{\Sigma_\tau} H +\int_{N_{[\tau,\tau_0]}} H\right) = 
\int_{\Sigma_\tau} 
\left( {\dot \pi} \delta \psi - {\dot \psi} \delta \pi \right)
+\int_{N_{[\tau,\tau_0]}} 
\left( {\dot \pi} \delta \psi - {\dot \psi} \delta \pi \right) +
\int_{\partial \Sigma_\tau} \pi^3 \delta \psi +\int_{\partial
N_{[\tau,\tau_0]}} \pi^{\overline u} \delta \psi \]
\normalsize\noindent
where the density of the Hamiltonian on $N$ is defined by
$H:=\pi\psi_{,u}$. The motivation for this choice of $H$ on $N$ is given
in \cite{APP}. Roughly speaking, it is a consequence of
the same universal formula (\ref{Lpole}) but integrated over a null surface.

The following relations (see \cite{APP})   
 \[ \left.
\pi^3\right|_{\partial \Sigma}=-\sin\theta\dot\psi =
\pi^{\overline u}\, ,\quad \partial \Sigma_\tau=S_\tau \, , \; \partial
N_{[\tau,\tau_0]} =S_\tau\cup S_{\tau_0}
\, , \; \partial\left(\Sigma_\tau\cup
N_{[\tau,\tau_0]}\right) = S_{\tau_0} \]
enable one to obtain the following variational formula
 \begin{equation}
\label{MB} - \delta m_{TB} = \int_{\Sigma_\tau\cup
N_{[\tau,\tau_0]}} \left( {\dot \pi}
\delta \psi - {\dot \psi} \delta \pi \right) + \int_{S_{\tau_0}} \pi
\delta \psi \end{equation} where $m_{TB} := \int_{\Sigma_\tau\cup
N_{[\tau,\tau_0]}} H =\int_{\Sigma_{\tau_0}} H $ is the TB
energy at retarded time $\tau_0$. 
Killing the term at $S_{\tau_0}$ in (\ref{MB})
 by an appropriate choice of the boundary conditions, our system becomes
Hamiltonian as a usual infinite dimensional dynamical system.  This
can be achieved, assuming for example that
 \[ \delta\psi |_{S_{\tau_0}} =0 \] 
 which simply means that $\psi$ is fixed at the time $\tau_0$.
  The precise meaning of those heuristic considerations will be given in
\cite{ptjjk}. Let us only stress that the quantity $m_{TB}$ may be rendered
unambiguous by adding the requirement that it cannot increase in retarded
time. This particular result has been shown in \cite{cjm}.

\subsection{\hspace{-5mm}.\hspace{2mm}Energy-Momentum Tensor and
Non-conservation Laws} 
 Let us consider the
energy-momentum tensor for the scalar field $\phi$ \[
T^\mu{_\nu} = \frac1{\sqrt{-\eta}}\left( p^\mu\phi_\nu
-\delta^\mu{_\nu} L \right) \] where $\eta:=\det\eta_{\mu\nu}$
and by $\delta^\mu{_\nu}$ we have denoted the Kronecker's delta.
For the Lagrangian $L$ describing scalar field $\phi$ the
canonical energy momentum is symmetric.  
From N\"other theorem
we have \[ \partial_\mu \left( \sqrt{-\eta} T^\mu{_\nu}X^\nu
\right)=0 \] for a Killing vector field $X^\mu$, and integrating
the above formula we obtain \be\label{d0T} \partial_0 \int_\Sigma
\sqrt{-\eta} T^0{_\nu}X^\nu =-\int_{\partial\Sigma} \sqrt{-\eta}
T^3{_\nu}X^\nu \ee

 Usually, when $\Sigma$ is a space-like surface
with the end at spatial infinity, the boundary term on the
right-hand side vanishes and the equation (\ref{d0T}) expresses
conservation law for the appropriate generator related to the
vector field $X$.  On the contrary, for the hyperboloid the
right-hand side does not vanish and (\ref{d0T}) expresses {\em
non-conservation} law.  It can be easily verified that for the
energy and angular momentum we have respectively 
\be\label{T00} \int_\Sigma
\sqrt{-\eta} T^0{_0} = \int_\Sigma H ={\cal H} \ee
 \be\label{Jz}
J_z:=\int_\Sigma\sqrt{-\eta} T^0{_\varphi} =\int_\Sigma \pi \psi_{,\varphi}
\ee
Moreover, for the energy ($X^\mu=\delta^\mu_0$) the boundary term on the
right-hand side of (\ref{d0T})  can be expressed in terms of $\dot\psi$
 \be\label{T30}
-\int_{\partial\Sigma} T^3{_0}\rho^{-4}\sin\theta\rd\theta\rd\varphi
 = \int_{\partial\Sigma\subset\scri^+}
\dot\psi^2\sin\theta\rd\theta\rd\varphi  \ee
Equalities (\ref{T00}) and (\ref{T30}) show obviously the equivalence
between (\ref{d0T}) and (\ref{Hdot}). 
This way we obtain the following result:
\begin{Theorem}
For massless scalar field the energy loss formulae in Hamiltonian form
(\ref{Hdot}) and in N\"other form (\ref{d0T}) are equivalent.
\end{Theorem}
In \cite{APP} we have shown that the same is true in electrodynamics. In
gravity there is no energy-momentum tensor but the formula analogous to
(\ref{Hdot}) is also true. The equation (\ref{Hdot}) is a special case of the
 more general equality, which can be formulated for any Killing
 vector field $X=X^\mu\partial_\mu$ in the following form
  \be\label{dotHX} \partial_0
 \left( \int_\Sigma\sqrt{-\eta} T^0{_\nu}X^\nu \right) =
 -\int_{\partial \Sigma} \pi^3 X^A\psi_A +\pi^3 X^0 \dot\psi \ee
   One can check by a direct computation that
$X^3|_{\scri^+}$=0, which simply means that the 
 Killing field $X$ (related to Poincar\'e group)
is tangent to the future null infinity $\scri^+$.

 Previous equation (\ref{Hdot}) (in the form of (\ref{dotHX}))
  corresponds to the vector field
 $X_H:=\partial_0$.

  The vector field corresponding to the linear momentum in $z$
direction \[ X_P:=-\frac{\cos\theta}{\sqrt{1+\rho^2}}\partial_0
-\rho^2\cos\theta\partial_3 -\rho\sin\theta\partial_{\theta} \; ,
\quad X_P|_{\scri^+}=-\cos\theta\partial_0 \] gives also the loss
formula 
\be \label{pz0} -\partial_0 P_z = \int_{\partial \Sigma}
\pi^3 X_P^0 \dot\psi = -\int_{S(s,0)}\sin\theta\cos\theta
(\dot\psi)^2 \ee
 where $\displaystyle P_z:= \int_\Sigma
\sqrt{-\eta}T^0_\mu X_P^\mu$.

The angular momentum along $z$-axis is related to the vector field
$X_J=\partial_\varphi$, and (\ref{dotHX}) takes the form
\be \label{Jdot} -\partial_0 J_z = \int_{\partial \Sigma}
\pi^3 X_J^A \psi_A = \int_{S(s,0)}\sin\theta
 \dot\psi\psi_{,\varphi} \ee

Similarly, we can take a boost generator along $z$-axis \[ X_K:=
-\rho\sqrt{1+\rho^2}\cos\theta\partial_3
-\sqrt{1+\rho^2}\sin\theta\partial_{\theta} +sX_P \; , \quad
X_K|_{\scri^+}=sX_P|_{\scri^+}+\hat\partial_\varphi \] where
$\hat\partial_A:=\varepsilon_A{^B}\partial_B$, and the
corresponding particular form of the formula (\ref{dotHX}) 
is the following
 \be\label{boost0} -\partial_0 K_z =
\int_{\partial \Sigma}\pi^3 X_K^0\dot\psi + \pi^3 X_K^A\psi_A =
-s\partial_0 P_z -\int_{S(s,0)} \sin^2\theta \dot\psi \psi_\theta
\ee for $\displaystyle K_z:= \int_\Sigma T^0_\mu X_K^\mu$ or \[
-\partial_0 K_z +s\partial_0 P_z= \int_{S(s,0)} \sin\theta
\dot\psi\hat\partial_\varphi \psi \]

The equations (\ref{Hdot})  and
(\ref{pz0})--(\ref{boost0}) express the {\it non-conservation} laws for the
Poincar\'e group generators defined at null infinity.

\section{\hspace{-4mm}.\hspace{2mm}GENERAL RELATIVITY}

\subsection{\hspace{-5mm}.\hspace{2mm}Generating Formula for Einstein
Equations} 
\label{generating}

The variation of the Hilbert Lagrangian (see \cite{Kij-qlh})
 \begin{equation} L = \frac 1{16
\pi}
\sqrt{|g|} \ R \label{Hilbert} \end{equation} may be calculated as follows
\begin{equation} \delta L = \delta \left( \frac 1{16 \pi} \sqrt{|g|} \
g^{\mu\nu} \ R_{\mu\nu} \right) = - \frac 1{16 \pi} {\cal G}^{\mu\nu}
\delta g_{\mu\nu} + \frac 1{16 \pi} \sqrt{|g|} \ g^{\mu\nu} \delta
R_{\mu\nu}
\label{deltaR} \end{equation} where \begin{equation} {\cal G}^{\mu\nu} :=
\sqrt{|g|} \ (R^{\mu\nu} - \frac 12 g^{\mu\nu} R)   \end{equation}
It was proved in \cite{Kij-qlh} that the last term in (\ref{deltaR}) is a
boundary term (a complete divergence).  For this purpose we denote
 \begin{equation}
\label{defpiA} {\pi}^{\mu\nu} := \frac 1{16 \pi} \sqrt{|g|} \ g^{\mu\nu} \
 \quad \mbox{and} \quad
 A^{\lambda}_{\mu\nu} := {\Gamma}^{\lambda}_{\mu\nu} -
{\delta}^{\lambda}_{(\mu} {\Gamma}^{\kappa}_{\nu ) \kappa} \end{equation}
  We have  \begin{eqnarray}
\partial_\lambda A^{\lambda}_{\mu\nu} & = & \partial_\lambda
{\Gamma}^{\lambda}_{\mu\nu} -
\partial_{(\mu} {\Gamma}^{\lambda}_{\nu ) \lambda} = 
R_{\mu\nu} + A^{\lambda}_{\mu\sigma} A^{\sigma}_{\nu\lambda} - \frac
13 A^{\lambda}_{\mu\lambda} A^{\sigma}_{\nu\sigma}   \end{eqnarray}
Hence, we obtain an identity \begin{eqnarray} \partial_\lambda \left(
{\pi}^{\mu\nu}
\delta A^{\lambda}_{\mu\nu} \right) & = &
{\pi}^{\mu\nu} \delta R_{\mu\nu} + \left( \nabla_\lambda
{\pi}^{\mu\nu}
\right) \delta A^{\lambda}_{\mu\nu}   \end{eqnarray}
 Due to the metricity of
$\Gamma$ we have $\nabla_\lambda {\pi}^{\mu\nu} = 0$.  This way we obtain
\begin{equation} \label{pidR} {\pi}^{\mu\nu} \delta R_{\mu\nu} =
\partial_\lambda \left( {\pi}^{\mu\nu} \delta A^{\lambda}_{\mu\nu} \right) =
\partial_\kappa \left( {\pi}_{\lambda}^{\ \mu\nu\kappa} \delta
{\Gamma}^{\lambda}_{\mu\nu} \right)  \end{equation} where we denote
\begin{equation} {\pi}_{\lambda}^{\ \mu\nu\kappa} := {\pi}^{\mu\nu}
\delta^\kappa_\lambda - {\pi}^{\kappa ( \nu} \delta^{\mu )}_\lambda 
\end{equation} Inserting (\ref{pidR}) into (\ref{deltaR}) we have
\begin{equation} \label{deltaL-grav} \delta L = - \frac 1{16 \pi} {\cal
G}^{\mu\nu} \delta g_{\mu\nu} + \partial_\lambda \left( {\pi}^{\mu\nu}
\delta A^{\lambda}_{\mu\nu} \right)   \end{equation} We conclude that
Euler-Lagrange equations ${\cal G}^{\mu\nu} = 0$ are equivalent to the
following generating formula, analogous to the (\ref{Lpole})
 in field theory
  \begin{equation} \delta L =
\partial_\lambda \left( {\pi}^{\mu\nu} \delta A^{\lambda}_{\mu\nu} \right)
\label{dL=pidA} \end{equation} or, equivalently, \begin{equation} \delta L =
\partial_\kappa \left( {\pi}_{\lambda}^{\ \mu\nu\kappa} \delta
{\Gamma}^{\lambda}_{\mu\nu} \right) \label{dL=pidgamma} 
\end{equation} This formula is a starting point for the derivation of
canonical gravity.  Let us observe, that it is valid not only in the
present, purely metric, context but also in any variational formulation of
General Relativity.  For this purpose let us rewrite (\ref{deltaL-grav})
without using {\em a priori} the metricity condition $\nabla_\lambda
{\pi}^{\mu\nu} = 0$.  This way we obtain the following, universal formula
\begin{equation} \delta L = - \frac 1{16 \pi} {\cal G}^{\mu\nu} \delta
g_{\mu\nu} - \left( \nabla_\kappa {\pi}_{\lambda}^{\
\mu\nu\kappa} \right) \delta {\Gamma}^{\lambda}_{\mu\nu} + \partial_\kappa
\left( {\pi}_{\lambda}^{\ \mu\nu\kappa} \delta {\Gamma}^{\lambda}_{\mu\nu}
\right) \end{equation} It may be proved that, in this form, the formula remains
valid also in the metric-affine approach and in the purely-affine one
(\cite{Kij-old}).  In metric-affine formulation, the vanishing of
$\nabla_\lambda {\pi}^{\mu\nu}$ is not automatic:  it is a part of field
equations.  We see that, again, the entire field dynamics is equivalent to
(\ref{dL=pidgamma}).  Finally, in the purely-affine formulation of General
Relativity the Einstein equations are satisfied ``from the very
beginning'' whereas the metricity condition for the connection becomes the
dynamical equation.  We conclude that also in this case the entire
information about the field dynamics is contained in the generating formula
(\ref{dL=pidgamma}). The universality of (\ref{dL=pidgamma}) enables one
to apply the same tool for different regimes, namely spatial and null
infinity. 

Let $(P^{kl},g_{kl})$ be a standard Cauchy data in ADM form \cite{ADM},
\cite{MTW} on a space-like surface.
In \cite{MG-qlh} and \cite{Kij-qlh} the ADM aspect of the formula
(\ref{dL=pidgamma}) has
been developed. In particular, the following theorem has been proposed:
\begin{Theorem}
Dynamical Einstein equations are equivalent to the following formula:
\be
\int_{V} \dot{  P}^{kl} \delta g_{kl} -
   \dot{g}_{kl} \delta {  P}^{kl}  =
   \int_{\partial V} {\check{  Q}}^{AB}
 \delta g_{AB} 
- 2  \int_{\partial V}\left[ n^A \delta {  P}^3{_A} +
 \frac{N^3}{\sqrt{\hat{g}^{33}}} \delta \frac{{ 
P}^{33}}{\sqrt{\hat{g}^{33}}} + N  \delta (\lambda k) \right]   \label{nielin}
\ee
where $\displaystyle {\check{  Q}}_{AB}= N^3{  P}_{AB}- N{\lambda}k_{AB}+
(Nk - \frac{\hat{g}^{3l}}{\sqrt{\hat{g}^{33}}} N,{_l}) \lambda {g}_{AB} $,
 $N=(-g^{00})^{-\frac{1}{2}}$ is the lapse function,
 $N^k = \hat{g}^{kl} g_{0l}$ is the shift, $\lambda:=\sqrt{\det g_{AB}}$,
  $\breve{g}^{AB}$ is the 2-dimensional inverse to $g_{AB}$,
$n^A =\breve{g}^{AB}g_{0B}$ and
by $k_{AB}$ we have denoted extrinsic curvature of the 2-surface $\partial V$
embedded in $V$ ($k$ is its 2-dimensional trace).
\end{Theorem}
This is an example of a homogeneous generating formula on a space-like
hypersurface. It enables one to derive quasi-local Hamiltonians in General
Relativity (cf. \cite{Kij-qlh}). The boundary term in (\ref{nielin}) can
be also 
written in an equivalent form described by eq. (80) in \cite{Kij-qlh}.
We will show in the sequel that the same formula (\ref{dL=pidgamma})
possesses an equally important status at null infinity and leads to the similar
result.

\subsection{\hspace{-5mm}.\hspace{2mm}Metrics of Bondi-Sachs Type}
It has been proved in \cite{APP} that the generating formula
(\ref{dL=pidgamma}) is applicable to the situation considered in 
\cite{BVM}, \cite{Burg}, \cite{Sachs}, \cite{XIV}.
The curved
space-time $M$ equipped with a pseudoriemannian metric of the form
\be\label{gB} 
g_{\mu\nu}{\rm d}x^\mu{\rm d}x^\nu= -\frac Vr \E{2\beta}{\rm d}u^2 
-2\E{2\beta}{\rm d}u{\rm d}r + r^2 \gamma_{AB}({\rm d}x^A -U^A {\rm
d}u)({\rm d}x^B -U^B {\rm d}u)
\ee
enables one to  consider the initial value problem on a
light-like hypersurface 
 \[ C:= \left\{ x\in M \;| \; x^0=u=\mbox{const.}\, ,\; r\geq r_0 \right\} \]
 with the boundary $\partial C = S(u,r=r_0)\cup S(u,r=\infty)$,
 where 
 \[ S(u,r=r_0):=\{ x \in M \; | \; x^0=u=\mbox{const.}\, ,\; r=r_0 \} \]
  and
 $S(u,r=\infty)=S_u\subset\scri^+$ is a sphere at the future null infinity.
We also assume that
\[ \sqrt{\det \gamma_{AB}}=\sin\theta \]

 The following boundary integral at null infinity, proposed by Trautman
and Bondi, defines the mass in the radiating regime
  \be\label{mB}
  m_{TB}:=\frac 1{8\pi} \int_{S_u} r-V = \frac 1{4\pi} \int_{S_u} M\sin\theta
  \ee
where $V=r-2M+O(r^{-1})$\footnote{The asymptotic behaviour
of the full metric $g_{\mu\nu}$ in the form (\ref{gB}) is given in
\cite{Burg} and summarized in \cite{APP}.}.
We would like to stress that in general the definition (\ref{mB}) is
correct only on a $\{ u=\mbox{const.} \}$ cross-section of $\scri^+$.
If we consider any (cross-) section $s: S^2\longrightarrow \scri^+$ of the
null infinity $\scri^+$, we can extend (\ref{mB}) to the following:\\
\underline{Definition}
\be 16\pi m_{TB}:= \int_{S^2}
\left(4M-{\kolo\chi}{^{AB}}{_{||AB}}\right)(s(\theta,\phi))\sin \theta
\rd\theta \rd\phi \label{dmTB}
\ee
\be 16\pi p^k:= \int_{S^2}
\left(4M-{\kolo\chi}{^{AB}}{_{||AB}}\right)(s(\theta,\phi))\frac{z^k}{r}
\sin \theta \rd\theta \rd\phi \label{dpTB}
\ee
 which enables one to prove the following theorem (cf. Section 8.1 in
\cite{APP}): 
 \begin{Theorem}
 The energy-momentum four-vector at null infinity is
invariant with respect to the passive supertranslations.
\end{Theorem}

Let us choose a (3+1)-foliation of space-time and integrate
(\ref{dL=pidA}) over a three-dimensional null-volume $V \subset C$ ($\partial
V =S_1\cup S_2$) 
\begin{equation} 
\delta \int_V L = \int_V \left( {\pi}^{\mu\nu} \delta
A^{0}_{\mu\nu} \right)\dot{} + \int_{\partial V} {\pi}^{\mu\nu} \delta
A^{3}_{\mu\nu}   \label{dLV}
\end{equation} 
 We use here adapted
coordinates; this means that the coordinate $x^3$ is constant on the boundary
$\partial V$.  Adapted coordinates simplify considerably derivation of the
final formula.  We stress, however, that all our results have an
independent, geometric meaning.  To rewrite them in a
coordinate-independent form it is sufficient to replace ``dots'' by Lie
derivatives ${\cal L}_X$, where $X$ is the vector field generating our
one-parameter group of transformations, which we are describing.  In
adapted coordinates $X:= \frac {\partial}{\partial x^0}$.  Moreover, the
upper index ``3'' has to be replaced everywhere by the sign ``$\perp$'',
denoting the transversal component with respect to the world tube.  This
way our results have a coordinate-independent meaning as relations between
well defined geometric objects and not just their specific components.

The following homogeneous generating formula can be directly obtained (cf.
\cite{APP}) from (\ref{dLV}) 
  \begin{eqnarray} 
 0 & = & \int_V \dot \Pi_{AB} \delta \psi^{AB} -
  \dot\psi^{AB} \delta \Pi_{AB}
  -\delta \int_{\partial V}2V\sin\theta+
 \nonumber \\ & &
+\frac 12 \int_{\partial V}\sin\theta \left(rV
{\gamma_{AB}}{_{,3}-r^2\dot{\gamma}_{AB} -2 U_{A||B}}
+r^2\E{-2\beta}U^C_{,3}\gamma_{CA}U_B \right)\delta
\gamma^{AB} + \nonumber \\ & & + \int_{\partial V} 2r^2\sin\theta
\left(\frac{2V}{r^2} -U^B{_{||B}} +U_A U^A_{,3}  \right) \delta\beta 
 - r^2\sin\theta \E{-2\beta} U_A \delta U^A_{,3}  
 \label{0nab} \end{eqnarray} 
 where we have introduced the following asymptotic variables
 $(\Pi_{AB},\psi^{AB})$ 
 analogous to the $(\pi,\psi)$ describing massless scalar field
\[ \psi^{AB}:=r\gamma^{AB} -r\kolo\gamma^{AB} \, , \quad
 \psi_{AB}:=r\gamma_{AB} -r\kolo\gamma_{AB}\]
\[ \Pi_{AB}:= -\frac 12 \sin\theta\left( r \gamma_{AB}\right)_{\! ,3}
 +\frac 12 \sin\theta\left( r \kolo\gamma_{AB}\right)_{\! ,3}\]
The equivalent form of the homogeneous formula (\ref{nielin}) expressed
 in terms of the natural geometric objects assigned to
the world tube $\{ x^3=\mbox{const.}\}$ ((80) in \cite{Kij-qlh})
 can be rewritten as follows
 \be\label{QonB} 
 0 =   \int_{V} \dot{ P}^{kl} \delta g_{kl} -
   \dot{g}_{kl} \delta {  P}^{kl}  +
 2\int_{\partial V} \dot\lambda\delta\alpha - \dot\alpha\delta\lambda +
  \int_{\partial V}
 \delta(n^2 Q^{00}) + n^2\delta Q^{00} -2 n^A\delta Q^0_A + Q^{AB}\delta
g_{AB} \ee
where $Q^{ab}$ is the tensor density built up from extrinsic curvature of the
(one-time-like-two-space-like) world tube in a similar way as ADM momentum
for space-like hypersurface,
$\lambda:=\sqrt{\det g_{AB}}$, $\alpha:=\arcsinh {g^{30}\over
\sqrt{-g^{00}g^{33}}}$, $n^A:=\breve{g}^{AB}g_{0B}$ and
$n:=\sqrt{n_An^A-g_{00}}$. 

The equation (\ref{QonB}) enables one to compose generating
formulae on a space-like and null 
hypersurfaces along two-dimensional boundary, where they meet. 
More precisely, let $O$ be a space-like hypersurface with $\partial
O=S(u,r_0)$, so $O\cup C$ gives a typical example of such composition. 
 One can directly check in Bondi coordinates that the
boundary terms on a sphere $S(u,r_0)$ in (\ref{QonB})  and
(\ref{0nab})  corresponding respectively to $O$ and $C$  are exactly
the same.  
This way we can write the variational formula on a 
truncated cone $O\cup C$, which is space-like inside and light-like near Scri.
One can also take a space-like hyperboloidal hypersurface $\Sigma_u$, which
approaches $\scri^+$ in an appropriate way by moving the ``sticking
together'' sphere $S(u,r_0)$ to the null infinity along cone $C$.
The above observations confirm the fact that TB mass is not sensitive
on the particular choice of the internal shape of the hypersurface but
depends only on its boundary, which is a section of $\scri^+$ (cf.
Definition of $m_{TB}$ before Theorem 3.2).

This way, passing to the limit\footnote{This makes sense even for a
polyhomogeneous asymptotics considered in \cite{XIV}.}, the composition of
the formula (\ref{0nab}) together with (\ref{QonB}) takes the form 
\be\label{Mham}
-16\pi\delta m_{TB}= \int_{O} \dot{  P}^{kl} \delta g_{kl} -
   \dot{g}_{kl} \delta {  P}^{kl}  +
 \int_C  \dot \Pi_{AB} \delta \psi^{AB} -
  \dot\psi^{AB} \delta \Pi_{AB}
 -\frac 12 \int_{S_u}\sin\theta \dot{\psi}_{AB} \delta
\psi^{AB} \ee

Similarly to (\ref{Hdot}),
one can denote the non-conservation law for the TB mass as follows
\be\label{M0}
-16\pi\partial_0 m_{TB}=
 -\frac 12 \int_{S_u}\sin\theta \dot{\psi}_{AB} 
\dot\psi^{AB} \; \left(=\frac 12 \int_{S_u}\sin\theta
\kolo\chi_{AB,u} \kolo\chi^{AB}{_{,u}} \right)
\ee 
where the last form in the brackets becomes clear when we apply the
asymptotics presented in \cite{APP}. In particular,
$\psi_{AB}|_{\scri^+}= \kolo\chi_{AB}$ and $\psi^{AB}|_{\scri^+}=
-\kolo\chi^{AB}$. 

The formula (\ref{M0}) is an example of the non-conservation law similar
to (\ref{Hdot}). It expresses the central result of the classical paper
\cite{BVM} and is valid in the form (\ref{M0}) for much wider asymptotics than
considered in the original papers \cite{BVM}, \cite{Burg}, \cite{Sachs}.
The property described by this law, namely
monotonicity in time for all vacuum field configurations, leads to the
uniqueness property of the TB energy, which we summarize in the
last section.\\
\underline{Remark.}
Quasi-local Hamiltonian $\displaystyle \frac1{16\pi}\int_{\partial V} n^2
Q^{00}$ renormalized by an additive constant gives simultaneously TB mass
and ADM mass, depending on the limit we take.

\section{\hspace{-4mm}.\hspace{2mm}UNIQUENESS OF THE TRAUTMAN-BONDI MASS}
 In  \cite{cjm} it is shown that 
 the TB energy is, up to a multiplicative constant $\alpha \in \Reals$, the
 only functional of the gravitational field, in a certain natural
class of functionals, which is  monotonic in time for all vacuum
  field configurations which admit (a piece of) a smooth null
infinity $\scri^+$. More precisely,  it is shown the
following:
\begin{Theorem}
\label{T.2}
Let $H$ be a functional of the form
\begin{equation}
\label{E3.1}
H[g,\,u] = 
\int_{S^2(u)} H^{\alpha\beta}(g_{\mu\nu},\,
g_{\mu\nu,\sigma},\,\ldots,\,g_{\mu\nu,\sigma_1\ldots\sigma_k}) \,\dx
S_{\alpha\beta},
\end{equation}
where the $H^{\alpha\beta}$ are twice differentiable functions of
their arguments, and the integral over ${S^2(u)}$ is understood as a
limit as $r_0$ goes to infinity of integrals over the spheres
$t=u+r_0,r=r_0$. Suppose that $H$ is monotonic in $u$ for all vacuum
metrics $g_{\mu\nu}$ for which $H$ is finite, provided that
$g_{\mu\nu}$ satisfies
\begin{eqnarray}&
{g_{\mu\nu} = \eta_{\mu\nu} +
\frac{h^1_{\mu\nu}(u,\,\theta,\,\varphi)}{r} +
\frac{h^2_{\mu\nu}(u,\,\theta,\,\varphi)}{r^2} + o(r^{-2})}\ ,
& \nonumber \\
&{\partial_{\sigma_1} \ldots \partial_{\sigma_i}( g_{\mu\nu} -
\frac{h^1_{\mu\nu}(u,\,\theta,\,\varphi)}{r} -
\frac{h^2_{\mu\nu}(u,\,\theta,\,\varphi)}{r^2}) = o(r^{-2}),\quad
}
\label{E.3.1}
\end{eqnarray}
with $ 1\leq i \leq k$, for some $C^k$ functions
$h^a_{\mu\nu}(u,\,\theta,\,\varphi)$, $a=1,2$. 
If $H$ is invariant under passive BMS 
super-translations, then the numerical value of $H$ equals (up to a
proportionality constant) the Trautman--Bondi mass.
\end{Theorem}
 Theorem
\ref{T.2} imposes the further requirement of {\em passive BMS
  invariance}, which did not occur in the scalar field case.
   We note that we believe that the requirement of
monotonicity  forces the energy to be invariant under (passive)
super--translations, but we have not succeeded in proving this so far.
The proof of Theorem \ref{T.2} is similar to the proof for the scalar
field, but technically rather more involved. A key
ingredient of the proof is the Friedrich-Kannar \cite{HF}, \cite{K}
construction of space-times ``having a piece of $\scri\,$''.
Let us finally mention that one can set up a Hamiltonian framework in
a phase space, which consists of Cauchy data on hyperboloids together
with values of the fields on appropriate parts of Scri, to describe the
dynamics in the radiation regime \cite{ptjjk}. Unsurprisingly, the
Hamiltonians one obtains in such a formalism are again not unique, but
the non-uniqueness can be controlled in a very precise way. The
Trautman-Bondi mass turns out to be a Hamiltonian, and an appropriate
version of the uniqueness Theorem \ref{T.2} can be used to single out
the TB mass amongst the family of all possible Hamiltonians. In the
Hamiltonian framework the freedom of multiplying the functional by a
constant disappears.


\begin{thebibliography}{666}
\footnotesize

\bibitem{BVM} 
H.~Bondi, M.G.J. van~der Burg, and A.W.K. Metzner,
   Proc. Roy. Soc. London A \textbf{269} (1962), 21-52.

\bibitem{Burg} 
M.G.J. van~der Burg, 
  Proc. Roy. Soc. London A \textbf{294} (1966),
  112-122.

\bibitem{Sachs} 
R.~Sachs, 
Phys. Rev.  \textbf{128} (1962), 2851-2864;
  Proc. Roy. Soc. London A \textbf{270}
  (1962), 103-126.
             
\bibitem{ADM} R. Arnowitt, S. Deser, C. Misner,
{\it The dynamics of general relativity}, in: Gravitation: an introduction
to current research, ed. L. Witten, p. 227 (Wiley, New York, 1962)

\bibitem{MTW} C. Misner, K.S. Thorne, J. A. Wheeler {\it Gravitation}, 
(W.H. Freeman and Co., San Francisco, 1973) 

\bibitem{BS} N. N. Bogoliubov, D. Shirkov, {\it Introduction into
Theory of Quantum Fields} (Science, Moscow, 1976)

\bibitem{Kij-Tulcz} J.  Kijowski and W.M.  Tulczyjew, {\it A Symplectic
Framework for Field Theories}, Lecture Notes in Physics No.107,
(Springer-Verlag, Berlin, 1979)

\bibitem{affine} 
J.~Kijowski, 
  Gen. Rel. Grav. \textbf{9} (1978),
  857-881.

\bibitem{Kij-old} M.~Ferraris, J.~Kijowski, Letters in Math. Phys.
\textbf{5} (1981), 127-135; Gen. Rel. Grav. \textbf{14} (1982), 165-180.

\bibitem{MG-qlh} J.~Jezierski, J.~Kijowski, 
{\em Quasi-local hamiltonian of the gravitational field}, in: Proceedings
of the Sixth Marcel Grossmann Meeting on General Relativity, Kyoto 1991,
Part A, editors H.  Sato and T.  Nakamura, p. 123-125, (World Scientific, 1992)

\bibitem{Kij-qlh} 
J.~Kijowski, 
  Gen. Rel. Grav. \textbf{29} (1997),
  307-343.

\bibitem{APP} J.~Jezierski, 
Acta Physica Polonica B 29 (1998) p. 667-743

\bibitem{XIV}
P.T. Chru\'sciel, M.A.H. MacCallum, and D.B. Singleton, 
  Phil. Trans. Roy. Soc. A \textbf{350} (1995),
  113-141.

\bibitem{cjm} P. Chru\'sciel, J. Jezierski and  M. MacCallum, 
Physical Review Letters 80 (1998) p. 5052-5055,
(gr-qc/9801073); 
 Physical Review D 58, 084001 (1998) 

\bibitem{ptjjk}
 P.T. Chru\'sciel, J. Jezierski, J. Kijowski, {\em A
Hamiltonian Framework for Field Theories in the Radiating Regime}, In
preparation 

\bibitem{HF}
H.~Friedrich, 
  Proc. Roy. Soc. London A \textbf{378} (1981), 401-421;
  Proc. Roy. Soc. London A
  \textbf{375} (1981), 169-184.

\bibitem{K}
J.~Kannar, 
  Proc. Roy. Soc.
  London A \textbf{452} (1996), 945-952.

\bibitem{Trautman}
A.~Trautman, \emph{King's {C}ollege lecture notes on general relativity},
  mimeographed notes (unpublished), May-June 1958;
  Bull. Acad. Pol. Sci., S\'erie sci. math., astr. et phys. \textbf{VI} (1958),
  407-412;
 \emph{{Conservation laws in general relativity}}, {Gravitation. An
  introduction to current research} ({Witten, L.}, ed.), {John Wiley and Sons},
  {New York and London}, {1962}.

\end{thebibliography}
\end{document}